\documentclass{ifacconf}
\pdfoutput=1
\usepackage{graphicx}      
\usepackage{natbib}        
\usepackage{url}
\usepackage{mathtools} 
\usepackage{amssymb}
\let\classtheoremstyle\theoremstyle
\let\theoremstyle\relax
\usepackage{amsthm}

\let\theoremstyle\classtheoremstyle
\usepackage{flushend}
\usepackage{balance} 
\usepackage{subcaption}
\usepackage{comment}
\usepackage{bbm}
\usepackage{tikz-cd}
\usepackage{nicefrac}
\usepackage{algorithm}
\usepackage{enumitem}

\let\classAND\AND
\let\AND\relax
\usepackage{algorithmic}

\let\AND\classAND
\AtBeginEnvironment{algorithmic}{\let\AND\algoAND}

\newtheorem{theorem}{Theorem}
\newtheorem{corollary}{Corollary}
\newtheorem{lemma}{Lemma}

\newtheorem{remark}{Remark}

\allowdisplaybreaks

\usepackage{environ}
\newcounter{quote}
\NewEnviron{myquote}
{%
\par\noindent\refstepcounter{quote}%
\parbox{0.9\linewidth}{%
\begin{quote}\BODY%
\end{quote}%
}\hfill\parbox{.1\linewidth}{(Q)}} 

\newcounter{assumpn}
\newenvironment{assumpn}[1][]{\refstepcounter{assumpn}
   \textbf{Assumption~\theassumpn:#1} \rmfamily}{}
\begin{document}
\begin{frontmatter}

\title{Consensus ADMM-Based Distributed Simultaneous Imaging \& Communication\thanksref{footnoteinfo}} 

\thanks[footnoteinfo]{This work was partially supported by the NSF Grant CNS-1956297, and the Sustainable Futures Fund \#919027. \\ © 2022 the authors. This work has been accepted to IFAC for publication under a Creative Commons Licence CC-BY-NC-ND.}

\author{Nishant Mehrotra} 
\author{Ashutosh Sabharwal}
\author{C\'esar A. Uribe} 

\address{Department
of Electrical and Computer Engineering, Rice University, Houston,
TX, 77005, USA (e-mail: \{nm30,~ashu,~cauribe\}@rice.edu).}

\begin{abstract}                
This paper takes the first steps toward enabling wireless networks to perform both imaging and communication in a distributed manner. We propose \emph{Distributed Simultaneous Imaging and Symbol Detection} (DSISD), a provably convergent distributed simultaneous imaging and communication scheme based on the alternating direction method of multipliers. We show that DSISD achieves similar imaging and communication performance as centralized schemes, with order-wise reduction in \text{computational} complexity. We evaluate the performance of DSISD via $2.4$ GHz Wi-Fi simulations.
\end{abstract}


\end{frontmatter}

\section{Introduction}
\label{sec:introduction}


In recent years, there has been growing interest in performing \emph{imaging} with wireless networks usually tailored for communications~\citep{Nanzer2019,Sadhu2021}. Wireless networks are ubiquitous and scalable by design, with powerful backend computing capabilities. This makes it possible to perform imaging alongside communication on a \emph{network-wide} scale.

\begin{figure}
    \centering
    \includegraphics[page=1,width=\linewidth]{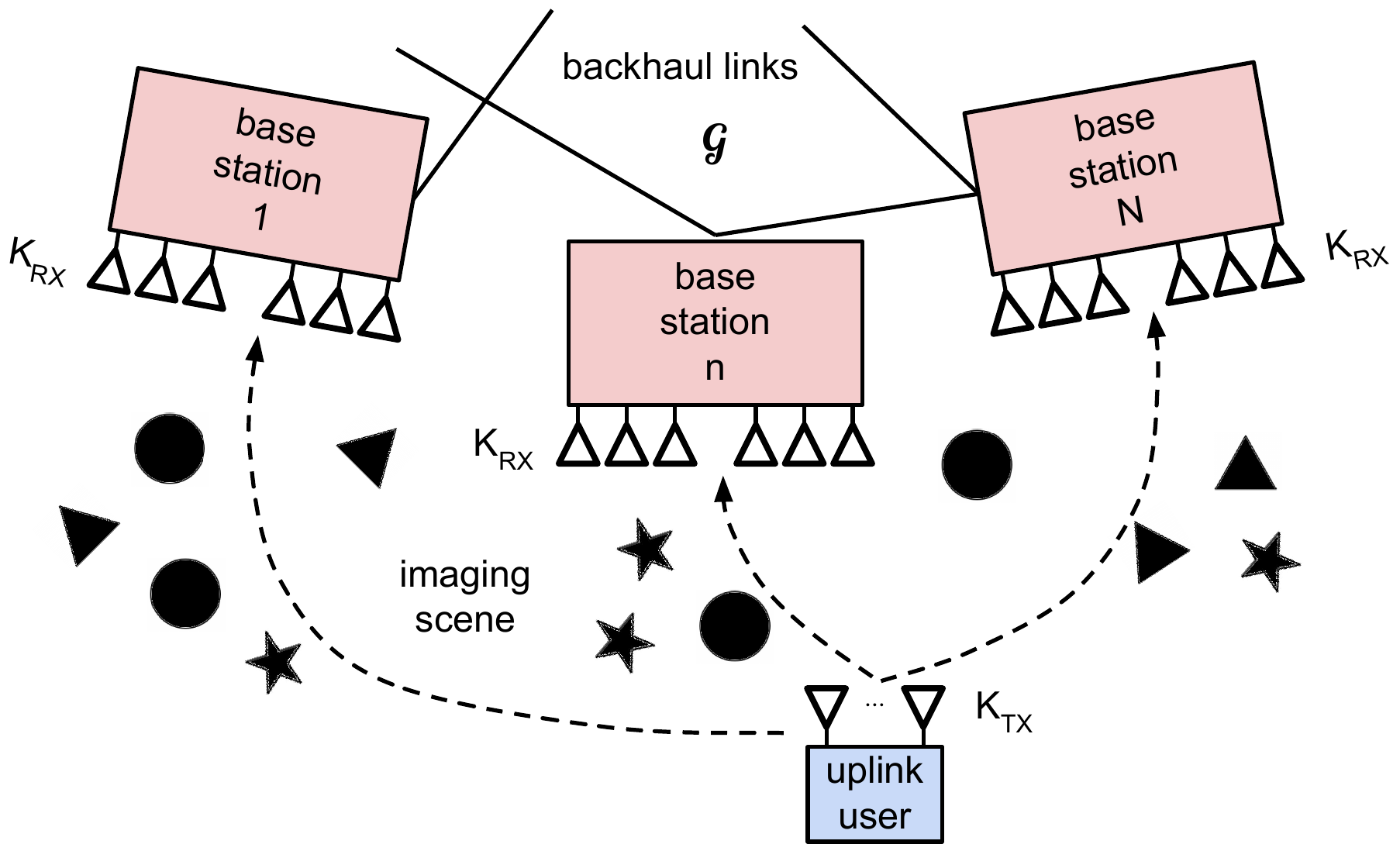}
    \caption{System configuration with $N$ base stations receiving transmissions from an uplink user. Base stations are connected over static, undirected backhaul network $\mathcal{G}$. Base stations aim to collaboratively image the scene and perform uplink user symbol detection.}
    \label{fig:setup}
\end{figure}

We study the problem of distributed simultaneous imaging and communication. We consider the system configuration shown in Fig.~\ref{fig:setup}, with $N \geq 1$ multiple-input multiple-output (MIMO) base stations receiving transmissions from an uplink user. The base stations aim to: (i) perform uplink user symbol detection (communication) and (ii) estimate the reflective response of scatterers in the environment (imaging). We assume the base stations can cooperate over a static, undirected graph $\mathcal{G}$, which models limited capacity backhaul links between the base stations. Due to the limited capacity of backhaul links, the base stations must perform the two operations (communication and imaging) in a distributed manner with local processing.

We propose \emph{Distributed Simultaneous Imaging and Symbol Detection} (DSISD), a provably convergent distributed algorithm for simultaneous imaging and communication. DSISD is the distributed variant of decode-and-image, a centralized algorithm previously proposed in~\citep{Mehrotra2021b}. We show that DSISD achieves similar imaging and communication performance as decode-and-image, with $O(N)$ reduction in \text{computational} complexity. We numerically evaluate the performance of DSISD via $2.4$ GHz Wi-Fi simulations.


To the best of our knowledge, we are the first to consider the problem of distributed simultaneous imaging and communication. Prior work has largely focused on communication-only (no imaging) and imaging-only (a-priori known communication data) problems. For instance, distributed algorithms have been proposed for beamforming, symbol detection, and interference alignment~\citep{Rajawat2016,Tao2017,Studer2017}, and for radar imaging~\citep{Mavrychev2018,Ottersten2021}. In the absence of priors on the scatterers in the environment and uplink data symbols, imaging-only and communication-only problems are convex. Nevertheless, we show that simultaneous imaging and communication corresponds to a bi-convex problem. We show that DSISD converges to the stationary points of the bi-convex problem with sublinear rate and guarantees asymptotic consensus across all $N$ base stations.

This paper is organized as follows. In Section~\ref{sec:sys_model}, we present the system model and problem formulation for Fig.~\ref{fig:setup}. 
We present the proposed algorithm, DSISD, and associated theoretical results in Section~\ref{sec:results}. In Section~\ref{sec:num_eval}, we show numerical results and simulations. We conclude the paper in Section~\ref{sec:conclusion} with some directions for future work.



\section{System Model}
\label{sec:sys_model}

Consider the system shown in Fig.~\ref{fig:setup} with $N \geq 1$ base stations receiving uplink transmissions. We adopt the system model proposed in~\citep{Mehrotra2021b}, where a similar configuration with a single base station was analyzed. In the sequel, we shall collectively refer to the set of scatterers that reflect uplink transmissions to the base stations as the \emph{imaging scene}. We make the following assumptions about the system operation.

\begin{assumpn}
\label{assumpn:1} 
\begin{itemize}[leftmargin=0.5cm]
    \item Scatterers in the imaging scene remain static for $T$ symbol durations, i.e., coherence interval.
    \item The uplink signalling is uni-polarized, with operating wavelength $\lambda$.
    \item All $N$ base stations are equipped with equal number of receive antennas, $K_{\mathsf{RX}}$, and operate synchronously over the same set of time-frequency resources.
\end{itemize}
\end{assumpn}




Formally, let $K_{\mathsf{TX}}$ denote the number of transmit antennas at the uplink user, and let $M$ denote the number of scatterers in the imaging scene. We denote the reflective response of the scatterers by a \emph{scene reflectivity} vector ${f} \in \mathbb{C}^{M}$. Furthermore, let ${X}_{T}$ denote the $K_{\mathsf{TX}} \times T$ matrix of transmitted uplink symbols, and ${N}_{T}^{(n)}$ denote the $K_{\mathsf{RX}} \times T$ matrix of additive noise at the $n$-th base station, for all $n \in \left\{1,\cdots,N\right\}$. Then, within a coherence interval of $T$ symbol durations, the $K_{\mathsf{RX}} \times T$ matrix of receive symbols at the $n$-th base station is given by
\begin{equation}
    \label{eq:sys1}
    {Y}_{T}^{(n)} = \underbrace{{P}_{\mathsf{RX}}^{(n)} \mathsf{diag}\big({f}\big) \big({P}_{\mathsf{TX}} \big)^{\top}}_{{H}_{\mathsf{comm}}^{(n)}} {X}_{T} + {N}_{T}^{(n)},
\end{equation}
where the two matrices ${P}_{\mathsf{TX}}$ and ${P}_{\mathsf{RX}}^{(n)}$ have sizes \mbox{$K_{\mathsf{TX}} \times M$} and \mbox{$K_{\mathsf{RX}} \times M$} respectively
The $\big(k,m\big)$-th element of each path delay matrix is a scaled complex exponential that depends on the signalling wavelength $\lambda$, and the locations of the $k$-th antenna and $m$-th scatterer in the scene,
\begin{equation*}
    {P}(k,m) = \frac{\exp({-j \big(\frac{2\pi}{\lambda}\big) \left\| {r}_{k} - {\tilde{r}}_{m} \right\|_{2}})}{4\pi \left\| {r}_{k} - {\tilde{r}}_{m} \right\|_{2}},~\forall {P} \in \{{P}_{\mathsf{TX}},{P}_{\mathsf{RX}}^{(n)}\},
\end{equation*}
where ${r}_{k}$ and ${\tilde{r}}_{m}$ denote the position vectors of the $k$-th antenna and $m$-th scatterer in the imaging scene.



In the next section, we formulate the distributed simultaneous imaging and communication problem. 

\section{Problem Formulation}
\label{sec:prob_formuln}

Given the system model in~(\ref{eq:sys1}), the base stations aim to collaboratively perform two functions:
\begin{enumerate}
    \item \textbf{Imaging:} Estimate the reflectivity vector ${f}$, and
    \item \textbf{Communication:} Estimate the uplink symbols ${X}_{T}$.
\end{enumerate}

Our goal is to enable both functionalities with only \emph{local} processing of the received symbols ${Y}_{T}^{(n)}$ at every base station $n \in \left\{1,\cdots,N\right\}$. Specifically, we make the following assumption on the prior knowledge at the base stations.

\begin{assumpn}
\label{assumpn:4} 
\end{assumpn}
The $n$-th base station has \emph{local} knowledge of received symbols ${Y}_{T}^{(n)}$, path delay matrices $\{{P}_{\mathsf{TX}},~{P}_{\mathsf{RX}}^{(n)}\}$ and discrete set $\mathcal{X}_{T-T_{1}}$ from which uplink data symbols ${X}_{T-T_{1}}$ are drawn. Examples of $\mathcal{X}_{T}$ are $\mathcal{X}_{T} = \{\pm 1\}^{K_{\mathsf{TX}} \times T}$ and $\mathcal{X}_{T} = \{(\pm 1 \pm j)/\sqrt{2}\}^{K_{\mathsf{TX}} \times T}$ for binary and quadrature phase-shift keying (BPSK and QPSK).

We assume the first $T_{1}$ symbols in ${X}_{T} = \big[{X}_{T_{1}}, {X}_{T-T_{1}}\big]$ to be pilot symbols \emph{known} to all $N$ base stations. The base stations thus aim to collaboratively solve the problem:
\begin{equation}
    \label{prb:1}
    \resizebox{.87\hsize}{!}{$\min\limits_{{X}_{T{-}T_{1}} \in \mathcal{X}_{T-T_{1}},{f}} \frac{1}{2} \sum\limits_{n = 1}^{N} \left\|{Y}_{T}^{(n)} {-} {P}_{\mathsf{RX}}^{(n)} \mathsf{diag}\big({f}\big) \big({P}_{\mathsf{TX}} \big)^{\top} {X}_{T} \right\|_{F}^{2}$}.
    \tag{P1}
\end{equation}

Problem~\eqref{prb:1} is an \emph{integer constrained bi-convex} problem, since the objective function is non-convex and non-separable in ${f}$ and ${X}_{T-T_{1}}$, but convex in each variable separately. We simplify the problem by relaxing the integer constraints, i.e., solve the unconstrained problem.

To formulate the distributed version of~\eqref{prb:1}, we assume the base stations are connected over a backhaul network, modeled as a graph $\mathcal{G}$.

\begin{assumpn}
\label{assumpn:5}
The graph $\mathcal{G} = \big(\mathcal{V},\mathcal{E}\big)$ is static and undirected, and every node knows who its neighbors are. 
\end{assumpn}

The distributed problem is formulated by defining local variables ${f}^{(n)}$ and ${X}_{T-T_{1}}^{(n)}$, with the constraint that local solutions of nodes connected by an edge in $\mathcal{E}$ are identical,
\begin{align}    
    \label{prb:2}
    &\min_{{X}_{T-T_{1}}^{(n)},{f}^{(n)}} \resizebox{.75\hsize}{!}{$\frac{1}{2} \sum\limits_{n = 1}^{N}\left\|{Y}_{T}^{(n)} {-} {P}_{\mathsf{RX}}^{(n)} \mathsf{diag}\big({f}^{(n)}\big) \big({P}_{\mathsf{TX}} \big)^{\top} {X}_{T}^{(n)} \right\|_{F}^{2}$} \nonumber \\ 
    &\text{s.t.}~{X}_{T-T_{1}}^{(i)} = {X}_{T-T_{1}}^{(j)},~{f}^{(i)} = {f}^{(j)},~\forall \big(i,j\big) \in \mathcal{E}.
    \tag{P2}
\end{align}


After solving~\eqref{prb:2}, nodes perform symbol detection via zero-forcing~\citep{Tse2005},
\begin{equation*}
    \hat{{X}}_{T-T_{1}}^{(n)} = \Pi_{\mathcal{X}_{T-T_{1}}}\big({X}_{T-T_{1}}^{(n)}\big),~\forall n \in \mathcal{V},
\end{equation*}
where $\Pi_{\mathcal{S}}\big(\cdot\big)$ denotes projection onto set $\mathcal{S}$.

\section{Main Results}
\label{sec:results}

We begin by analyzing first-order optimality conditions in Section~\ref{subsec:results_opt_condn} to characterize the number of uplink pilots~$T_{1}$ required to solve~\eqref{prb:1}. We subsequently present the proposed DSISD algorithm in Section~\ref{subsec:results_algo}, and derive associated convergence guarantees in Section~\ref{subsec:results_algo_perf}.

\subsection{Optimality Conditions for Problem~\eqref{prb:1}}
\label{subsec:results_opt_condn}



Let the objective function in~\eqref{prb:1} be denoted by
\begin{equation*}
    \mathcal{L}\big({f},{X}_{T-T_{1}}\big) = \frac{1}{2} \sum\limits_{n = 1}^{N} \big\|{Y}_{T}^{(n)} {-} {P}_{\mathsf{RX}}^{(n)} \mathsf{diag}\big({f}\big) \big({P}_{\mathsf{TX}} \big)^{\top} {X}_{T} \big\|_{F}^{2}. 
\end{equation*}

The KKT conditions for~\eqref{prb:1} correspond to
\begin{equation*}
    \nabla_{{f}} \mathcal{L}\big({f}^{*},{X}_{T-T_{1}}^{*} \big) = {0},~\nabla_{{X}_{T-T_{1}}} \mathcal{L}\big({f}^{*},{X}_{T-T_{1}}^{*} \big) = {0},
\end{equation*}
where ${f}^{*}$ and ${X}_{T-T_{1}}^{*}$ are minimizers of~\eqref{prb:1}. Hence, the stationary points of~\eqref{prb:1} are given by
\begin{align}
    \label{eq:opt3}
    {f}^{*} &= \big(\left.{H}_{\mathsf{img}}\right|_{{X}_{T-T_{1}} = {X}_{T-T_{1}}^{*}}\big)^{\dagger} \mathsf{vec}\big({Y}_{T}\big), \\
    \label{eq:opt4}
    {X}_{T-T_{1}}^{*} &= \big(\left.{H}_{\mathsf{comm}}\right|_{{f} = {f}^{*}}\big)^{\dagger} {Y}_{T-T_{1}},
\end{align}
where $(\cdot)^{\dagger}$ denotes the pseudo-inverse and $\mathsf{vec}(\cdot)$ denotes the vectorization operator. The matrix ${H}_{\mathsf{img}}$ is given by
\begin{equation*}
    {H}_{\mathsf{img}} = \big({X}_{T}^{\top} \otimes {I}_{N K_{\mathsf{RX}}} \big) \big({P}_{\mathsf{TX}} \ast {P}_{\mathsf{RX}} \big),
\end{equation*}
where $\otimes$ and $\ast$ denote the Kronecker and column-wise Khatri-Rao products, and ${I}_{n}$ denotes the identity matrix of size $n \times n$. Matrices ${P}_{\mathsf{RX}}$, ${H}_{\mathsf{comm}}$ and ${Y}_{T}$ are concatenations of local path delay, channel and received symbol matrices across all $N$ base stations.




\begin{algorithm}[tb]
    \caption{\emph{Distributed Simultaneous Imaging \& Symbol Detection} (DSISD)}
    \label{alg:admm_dist}
    \textbf{Input:}~${P}_{\mathsf{TX}}$,~${P}_{\mathsf{RX}}^{(n)}$,~${Y}_{T}^{(n)}$,~$\mathcal{X}_{T-T_{1}}$,~${X}_{T_{1}}$,~$K$,~$\rho_{X}$,~$\rho_{f}$
\begin{algorithmic}[1]
   \FOR{node $n=1$ {\bfseries to} $N$ in $\mathcal{V}$ \COMMENT{in parallel}}
   \STATE Initialize ${f}_{0}^{(n)} = {D}_{f,0}^{(n)} = {0},{X}_{T-T_{1},0}^{(n)} = {D}_{X,0}^{(n)} = {0}$
   \FOR{iteration $k=1$ {\bfseries to} $K$}
   \STATE Transmit ${f}_{k-1}^{(n)}$, ${X}_{T-T_{1},k-1}^{(n)}$ to $\big\{m: (n,m) \in \mathcal{E}\big\}$
   \STATE Receive $\big\{{f}_{k-1}^{(m)},~{X}_{T-T_{1},k-1}^{(m)},~\forall m: (n,m) \in \mathcal{E}\big\}$
   \STATE Update reflectivities ${f}_{k}^{(n)}$ by solving
   \begin{equation*}
        \min_{{f}} \left\{
        \begin{aligned}
            &\resizebox{.65\hsize}{!}{$\frac{\rho_{f}}{2} \big\| \sum\limits_{m} {L}_{\mathcal{G}}(n,m) {f}_{k-1}^{(m)} {+} {L}_{\mathcal{G}}(n,n) {f} {+} {D}_{f,k-1}^{(n)} \big\|_{2}^{2}$} \\ &\resizebox{.65\hsize}{!}{${+} \frac{1}{2} \big\|{Y}_{T}^{(n)} {-} {P}_{\mathsf{RX}}^{(n)} \mathsf{diag}\big({f}\big) \big({P}_{\mathsf{TX}} \big)^{\top} {X}_{T,k-1}^{(n)} \big\|_{F}^{2}$}
        \end{aligned}\right\}
   \end{equation*}
   \STATE Update uplink symbols ${X}_{T-T_{1},k}^{(n)}$ by solving
   \begin{equation*}
        \min_{{X}} \left\{
        \begin{aligned}
            &\resizebox{.65\hsize}{!}{$\frac{\rho_{X}}{2} \big\| \sum\limits_{m} {L}_{\mathcal{G}}(n,m) {X}_{T-T_{1},k-1}^{(m)} {+} {L}_{\mathcal{G}}(n,n) {X} {+} {D}_{X,k-1}^{(n)} \big\|_{2}^{2}$} \\ &\resizebox{.65\hsize}{!}{${+} \frac{1}{2} \big\|{Y}_{T}^{(n)} {-} {P}_{\mathsf{RX}}^{(n)} \mathsf{diag}\big({f}_{k}^{(n)}\big) \big({P}_{\mathsf{TX}} \big)^{\top} {X} \big\|_{F}^{2}$}
        \end{aligned}\right\}
   \end{equation*}
   \STATE Update dual variables via
   \begin{align*}
       &\resizebox{.75\hsize}{!}{${D}_{f,k}^{(n)} = {D}_{f,k-1}^{(n)} {+} \sum\limits_{m} {L}_{\mathcal{G}}(n,m) {f}_{k}^{(m)} {+} {L}_{\mathcal{G}}(n,n) {f}_{k}^{(n)}$} \\
       &\resizebox{.85\hsize}{!}{${D}_{X,k}^{(n)} = {D}_{X,k-1}^{(n)} {+} \sum\limits_{m} {L}_{\mathcal{G}}(n,m) {X}_{T-T_{1},k}^{(m)} {+} {L}_{\mathcal{G}}(n,n) {X}_{T-T_{1},k}^{(n)}$}
   \end{align*}
   \ENDFOR
   \ENDFOR
   \STATE \textbf{Output:} $\hat{{f}} = {f}_{K}^{(N)}$, $\hat{{X}}_{T-T_{1}} = \Pi_{\mathcal{X}_{T-T_{1}}}\big({X}_{T-T_{1},K}^{(N)}\big)$
\end{algorithmic}
\end{algorithm}

Note that the values of~\eqref{eq:opt3} and~\eqref{eq:opt4} are coupled. In the following lemma, we characterize conditions under which stationary points satisfying~(\ref{eq:opt3}) and~(\ref{eq:opt4}) are \emph{decoupled}.



\begin{lemma}
    \label{lmm:1}
    Let ${N}_{T} = {0}$. If ${X}_{T_{1}} \big({X}_{T_{1}}^{*}\big)^{\dagger} = {I}_{K_{\mathsf{TX}}}$, or equivalently, $T_{1} \geq K_{\mathsf{TX}}$ for time-orthogonal pilots, then~\eqref{eq:opt3} and~\eqref{eq:opt4} are decoupled.
\end{lemma}
\begin{proof}
In the absence of noise, ${f}^{*}$ from~\eqref{eq:opt3} is given by
\begin{equation*}
    \resizebox{.99\hsize}{!}{${f}^{*} = \big({P}_{\mathsf{TX}} \ast {P}_{\mathsf{RX}} \big)^{\dagger} \big(\big({X}_{T} \big({X}_{T}^{*}\big)^{\dagger}\big)^{\top} \otimes {I}_{N K_{\mathsf{RX}}} \big) \big({P}_{\mathsf{TX}} \ast {P}_{\mathsf{RX}} \big) {f}.$}
\end{equation*}

The above expression is decoupled with the recovered uplink symbols ${X}_{T}^{*}$ when ${X}_{T} \big({X}_{T}^{*}\big)^{\dagger} = {I}_{K_{\mathsf{TX}}}$. Since data portion of ${X}_{T}$ and recovered symbols ${X}_{T}^{*}$ may differ, a sufficient condition is thus ${X}_{T_{1}} \big({X}_{T_{1}}^{*}\big)^{\dagger} = {I}_{K_{\mathsf{TX}}}$. For time-orthogonal pilot symbols, this condition is equivalent to
\begin{equation*}
    \mathsf{rank}\big({X}_{T_{1}}\big) = \min\big\{T_{1},K_{\mathsf{TX}}\big\} \geq K_{\mathsf{TX}} \implies T_{1} \geq K_{\mathsf{TX}}.
\end{equation*}
\end{proof}


Following Lemma~\ref{lmm:1}, we assume $T_{1} = K_{\mathsf{TX}}$ in all subsequent discussion. We now present the proposed DSISD algorithm for solving~\eqref{prb:2}.


\subsection{Distributed Simultaneous Imaging \& Symbol Detection}
\label{subsec:results_algo}

The bi-convexity of the objective function $\mathcal{L}\big({f},{X}_{T-T_{1}}\big)$ naturally motivates using an \emph{alternating} procedure to solve~\eqref{prb:2}. The proposed DSISD algorithm (Algorithm~\ref{alg:admm_dist}) is thus based on consensus ADMM~\citep{Boyd2011}. 

Let the concatenation of local solutions across all $N$ nodes be denoted by variables \mbox{$\overline{{X}}^{\top} = \big[\big({X}_{T-T_{1}}^{(1)}\big)^{\top},\cdots,\big({X}_{T-T_{1}}^{(N)}\big)^{\top} \big]$} and \mbox{$\overline{{f}}^{\top} = \big[\big({f}^{(1)}\big)^{\top},\cdots,\big({f}^{(N)}\big)^{\top}\big]$}. Moreover, let
\begin{equation*}
    \resizebox{.99\hsize}{!}{$g\big(\overline{{X}},\overline{{f}}\big) = \frac{1}{2} \sum\limits_{n = 1}^{N}\big\|{Y}_{T}^{(n)} {-} {P}_{\mathsf{RX}}^{(n)} \mathsf{diag}\big({f}^{(n)}\big) \big({P}_{\mathsf{TX}} \big)^{\top} {X}_{T}^{(n)} \big\|_{F}^{2}$}
\end{equation*}
denote the objective function in~\eqref{prb:2}. Finally, \mbox{${L}_{\mathcal{G}} \in \mathbb{R}^{N \times N}$} denotes the Laplacian matrix for $\mathcal{G}$. 

With above defined notation,~\eqref{prb:2} may be recast as
\begin{align}
    \label{prb:3}
    &\min_{\overline{{X}},\overline{{f}}} g\big(\overline{{X}},\overline{{f}}\big) \nonumber \\
    &\text{s.t.}~\big({L}_{\mathcal{G}} \otimes {I}_{K_{\mathsf{TX}}} \big) \overline{{X}} = {0},~\big({L}_{\mathcal{G}} \otimes {I}_{M} \big)  \overline{{f}} = {0}.
    \tag{P3}
\end{align}

At every iteration $k \in \{0,1,\cdots\}$, DSISD performs the following ADMM updates:
\begin{align}
    \label{eq:alg1}
    \overline{{f}}^{k+1} &= \arg\min_{\overline{{f}}} \mathcal{A}\big(\overline{{X}}^{k},\overline{{f}},\overline{{D}}_{X}^{k},\overline{{D}}_{f}^{k}\big),
    \tag{A1} \\
    \label{eq:alg2}
    \overline{{X}}^{k+1} &= \resizebox{.55\hsize}{!}{$\arg\min\limits_{\overline{{X}}} \mathcal{A}\big(\overline{{X}},\overline{{f}}^{k+1},\overline{{D}}_{X}^{k},\overline{{D}}_{f}^{k}\big),$}
    \tag{A2} \\
    \label{eq:alg3}
    \overline{{D}}_{f}^{k+1} &= \overline{{D}}_{f}^{k} {+} \big({L}_{\mathcal{G}} \otimes {I}_{M} \big) \overline{{f}}^{k+1}.
    \tag{A3} \\
    \label{eq:alg4}
    \overline{{D}}_{X}^{k+1} &= \overline{{D}}_{X}^{k} {+} \big({L}_{\mathcal{G}} \otimes {I}_{K_{\mathsf{TX}}} \big) \overline{{X}}^{k+1},
    \tag{A4}
\end{align}
where $\mathcal{A}\big(\overline{{X}},\overline{{f}},\overline{{D}}_{X},\overline{{D}}_{f}\big)$ denotes the augmented Lagrangian,
\begin{multline*}
    \mathcal{A}\big(\overline{{X}},\overline{{f}},\overline{{D}}_{X},\overline{{D}}_{f}\big) = g\big(\overline{{X}},\overline{{f}}\big) {+} \frac{\rho_{f}}{2} \big\|\big({L}_{\mathcal{G}} \otimes {I}_{M} \big) \overline{{f}} \big\|_{F}^{2} \\
    {+} \rho_{X} \left\langle \overline{{D}}_{X}, \big({L}_{\mathcal{G}} \otimes {I}_{K_{\mathsf{TX}}} \big) \overline{{X}} \right\rangle {+} \rho_{f} \left\langle \overline{{D}}_{f}, \big({L}_{\mathcal{G}} \otimes {I}_{M} \big) \overline{{f}} \right\rangle \\ {+} \frac{\rho_{X}}{2} \big\|\big({L}_{\mathcal{G}} \otimes {I}_{K_{\mathsf{TX}}} \big) \overline{{X}} \big\|_{F}^{2}.
\end{multline*}

The variables $\overline{{D}}_{X}$ and $\overline{{D}}_{f}$ denote scaled dual variables, whereas $\rho_{X}$ and $\rho_{f}$ denote penalty parameters.

The updates in~(\ref{eq:alg1}) to~(\ref{eq:alg4}) can be performed locally at every base station since $g\big(\overline{{X}},\overline{{f}}\big)$ and $\mathcal{A}\big(\overline{{X}},\overline{{f}},\overline{{D}}_{X},\overline{{D}}_{f}\big)$ are separable across pairs of local solutions $\big({f}^{(n)},{X}_{T}^{(n)}\big)$. Below, we derive performance guarantees for DSISD. 


\subsection{Performance Guarantees for DSISD}
\label{subsec:results_algo_perf}

In Theorem~\ref{thm:1} and Corollary~\ref{corr:1}, we characterize convergence guarantees and algorithm complexity for DSISD. 
\\
\begin{theorem}
\label{thm:1}
Let Assumptions~\ref{assumpn:1},~\ref{assumpn:4},~\ref{assumpn:5} and the conditions in Lemma~\ref{lmm:1} hold. Then, the updates of DSISD satisfy the following properties:

\textbf{Asymptotic consensus:} All $N$ nodes in $\mathcal{G}$ are in consensus asymptotically, i.e.,
\begin{align*}
    \lim_{k \rightarrow \infty} \big\|\big({L}_{\mathcal{G}} \otimes {I}_{M} \big) \overline{{f}}^{k+1}\big\|_{2} &= 0, \\
    \lim_{k \rightarrow \infty} \big\|\big({L}_{\mathcal{G}} \otimes {I}_{K_{\mathsf{TX}}} \big) \overline{{X}}^{k+1}\big\|_{F} &= 0.  
\end{align*}

\textbf{Convergence to stationary points:} Limit points of iterates $\big(\overline{{X}}^{k},\overline{{f}}^{k},\overline{{D}}_{X}^{k},\overline{{D}}_{f}^{k} \big)$ converge to a KKT point of~\eqref{prb:3}.

\textbf{Sublinear convergence rate:} Iterates $\big(\overline{{X}}^{k},\overline{{f}}^{k},\overline{{D}}_{X}^{k},\overline{{D}}_{f}^{k} \big)$ converge to a KKT point of~\eqref{prb:3} with rate $O\big({1}/{K}\big)$ in terms of the optimality gap
\begin{multline*}
    Q\big(\overline{{X}}^{k+1},\overline{{f}}^{k+1},\overline{{D}}_{X}^{k},\overline{{D}}_{f}^{k} \big) = \big\|\big({L}_{\mathcal{G}} \otimes {I}_{K_{\mathsf{TX}}} \big) \overline{{X}}^{k+1}\big\|_{F}^{2} {+} \\ \big\|\nabla_{\overline{{f}}} \mathcal{A}\big(\overline{{X}}^{k+1},\overline{{f}}^{k+1},\overline{{D}}_{X}^{k},\overline{{D}}_{f}^{k} \big) \big\|_{F}^{2} {+} \big\|\big({L}_{\mathcal{G}} \otimes {I}_{M} \big) \overline{{f}}^{k+1}\big\|_{2}^{2} \\ {+} \big\|\nabla_{\overline{{X}}} \mathcal{A}\big(\overline{{X}}^{k+1},\overline{{f}}^{k+1},\overline{{D}}_{X}^{k},\overline{{D}}_{f}^{k} \big) \big\|_{F}^{2}.
\end{multline*}
\end{theorem}

\begin{proof}
Due to space constraints, we only provide a proof sketch with key steps and ideas for the complete proof.

\textbf{Result~$1$: Asymptotic Consensus}

We seek to prove
\begin{equation*}
    \lim_{k \rightarrow \infty} \big\|\overline{{D}}_{f}^{k+1}{-}\overline{{D}}_{f}^{k} \big\|_{2} = 0,~\lim_{k \rightarrow \infty} \big\|\overline{{D}}_{X}^{k+1}{-}\overline{{D}}_{X}^{k} \big\|_{2} = 0.
\end{equation*}

We show the above result via two intermediate steps.

\textbf{Step~$1$:} We first derive upper bounds on successive dual update norms, $\big\|\overline{{D}}_{f}^{k+1}{-}\overline{{D}}_{f}^{k} \big\|_{2}$ and $\big\|\overline{{D}}_{X}^{k+1}{-}\overline{{D}}_{X}^{k} \big\|_{2}$. Consider the first-order optimality condition for~\eqref{eq:alg2},
\begin{equation*}
    \nabla_{\overline{{X}}} g\big(\overline{{X}}^{k+1},\overline{{f}}^{k+1}\big) {+} \rho_{X} \big({L}_{\mathcal{G}}^{\top} \otimes {I}_{K_{\mathsf{TX}}} \big) \overline{{D}}_{X}^{k+1} = {0}.
\end{equation*}
Thus, an upper bound on $\big\|\overline{{D}}_{X}^{k+1} {-} \overline{{D}}_{X}^{k} \big\|_{F}^{2}$ is given by
\begin{equation*}
    \resizebox{.99\hsize}{!}{$\big\| \overline{{D}}_{X}^{k+1} {-} \overline{{D}}_{X}^{k} \big\|_{F}^{2} \leq \frac{\nicefrac{1}{\rho_{X}^{2}}}{\lambda_{\mathsf{min}}\big({L}_{\mathcal{G}}^{\top} {L}_{\mathcal{G}}\big)} \big\| \nabla_{\overline{{X}}} g\big(\overline{{X}}^{k},\overline{{f}}^{k}\big) {-} \nabla_{\overline{{X}}} g\big(\overline{{X}}^{k+1},\overline{{f}}^{k+1}\big) \big\|_{F}^{2}$},
\end{equation*}
where we have used the fact that $\frac{1}{2} \big\|\big({L}_{\mathcal{G}} \otimes {I}_{K_{\mathsf{TX}}} \big) \overline{{X}} \big\|_{F}^{2}$ is $\lambda_{\mathsf{min}}\big({L}_{\mathcal{G}}^{\top} {L}_{\mathcal{G}}\big)$-strongly convex.

On appropriate substitutions, the right hand side can be upper bounded in terms of successive \emph{primal} update norms, $\big\| \overline{{X}}^{k+1} {-} \overline{{X}}^{k} \big\|_{F}^{2}$, $\big\| \overline{{f}}^{k+1} {-} \overline{{f}}^{k} \big\|_{F}^{2}$. Thus, we next show $\lim_{k \rightarrow \infty} \big\|\overline{{f}}^{k+1} {-} \overline{{f}}^{k}\big\|_{2} = 0$ and $\lim_{k \rightarrow \infty} \big\|\overline{{X}}^{k+1} {-} \overline{{X}}^{k}\big\|_{F} = 0$.



\textbf{Step~$2$:} We equivalently show $\sum\limits_{k = 1}^{\infty} \big\| \overline{{X}}^{k+1} {-} \overline{{X}}^{k} \big\|_{F}^{2} < \infty$ and $\sum\limits_{k = 1}^{\infty} \big\| \overline{{f}}^{k+1} - \overline{{f}}^{k} \big\|_{2}^{2} < \infty$ via two sub-results. 

(i) First, we show that the augmented Lagrangian is upper bounded in terms of the primal update norms as
\begin{multline*}
    \mathcal{A}\big(\overline{{X}}^{k+1},\overline{{f}}^{k+1},\overline{{D}}_{X}^{k},\overline{{D}}_{f}^{k}\big) {-} \mathcal{A}\big(\overline{{X}}^{k},\overline{{f}}^{k},\overline{{D}}_{X}^{k},\overline{{D}}_{f}^{k}\big) \\ \resizebox{.99\hsize}{!}{$\leq {-}\frac{ \lambda_{\mathsf{min}}\big({L}_{\mathcal{G}}^{\top} {L}_{\mathcal{G}}\big)}{2} \big(\rho_{X} \big\| \overline{{X}}^{k+1} {-} \overline{{X}}^{k} \big\|_{F}^{2} {+} \rho_{f} \big\| \overline{{f}}^{k+1} {-} \overline{{f}}^{k} \big\|_{2}^{2} \big).$}
\end{multline*}

We show the above by upper bounding the left hand side and using the first-order optimality conditions corresponding to~\eqref{eq:alg1} and~\eqref{eq:alg2}, the convexity of the objective function $g\left(\overline{ {X}},\overline{ {f}}\right)$ in $\overline{ {X}}$ (resp. $\overline{ {f}}$) given fixed $\overline{ {f}}$ (resp. $\overline{ {X}}$), as well as the $\lambda_{\mathsf{min}}\big({L}_{\mathcal{G}}^{\top} {L}_{\mathcal{G}}\big)$ strong convexity of the terms $\frac{1}{2} \big\|\big({L}_{\mathcal{G}} \otimes {I}_{K_{\mathsf{TX}}} \big) \overline{{X}} \big\|_{F}^{2}$ and $\frac{1}{2} \big\|\big({L}_{\mathcal{G}} \otimes {I}_{K_{\mathsf{M}}} \big) \overline{{f}} \big\|_{2}^{2}$.

%

(ii) Next, we show that the augmented Lagrangian is lower bounded at every iteration, i.e.,
\begin{align*}
    &\sum\limits_{k = 1}^{K}  \mathcal{A}\big(\overline{{X}}^{k+1},\overline{{f}}^{k+1},\overline{{D}}_{X}^{k},\overline{{D}}_{f}^{k}\big) > -\infty, \\
    & \sum\limits_{k = 1}^{K} \mathcal{A}\big(\overline{{X}}^{k},\overline{{f}}^{k},\overline{{D}}_{X}^{k},\overline{{D}}_{f}^{k}\big) > -\infty.
\end{align*}

To show the above result, we utilize the dual updates in~\eqref{eq:alg3} and~\eqref{eq:alg4}, as well as the equality
\begin{equation*}
    \big\langle C,D \big\rangle = \frac{1}{2} \bigg(\big\|C+D \big\|_{F}^{2} - \big\|C \big\|_{F}^{2} - \big\|D \big\|_{F}^{2}\bigg),
\end{equation*}
for any arbitrary $C$ and $D$.

On taking the limit $K \rightarrow \infty$ in (i) and (ii) above, we obtain $\sum\limits_{k = 1}^{\infty} \big\| \overline{{X}}^{k+1} {-} \overline{{X}}^{k} \big\|_{F}^{2} < \infty$ and $\sum\limits_{k = 1}^{\infty} \big\| \overline{{f}}^{k+1} {-} \overline{{f}}^{k} \big\|_{2}^{2} < \infty$.

\textbf{Result~$2$: Convergence to Stationary Points}

The KKT conditions corresponding to~\eqref{prb:3} are
\begin{gather*}
    \nabla_{\overline{{X}}} g\big(\overline{{X}}^{*},\overline{{f}}^{*}\big) {+} \rho_{X} \big({L}_{\mathcal{G}}^{\top} \otimes {I}_{K_{\mathsf{TX}}} \big) \overline{{D}}_{X}^{*} = {0}, \\
    \nabla_{\overline{{f}}} g\big(\overline{{X}}^{*},\overline{{f}}^{*}\big) {+} \rho_{X} \big({L}_{\mathcal{G}}^{\top} \otimes {I}_{M} \big) \overline{{D}}_{f}^{*} = {0}, \\
    \big({L}_{\mathcal{G}} \otimes {I}_{K_{\mathsf{TX}}} \big) \overline{{X}}^{*} = {0},~\big({L}_{\mathcal{G}} \otimes {I}_{M} \big) \overline{{f}}^{*} = {0}.
\end{gather*}

We aim to show that the limit points corresponding to Algorithm~\ref{alg:admm_dist} satisfy the above KKT conditions. To that end, observe that in the limit $k \rightarrow \infty$, the update steps~(\ref{eq:alg1}) and~(\ref{eq:alg2}) satisfy
\begin{gather*}
    \resizebox{.99\hsize}{!}{$\nabla_{\overline{{X}}} \mathcal{A} = {0} \implies \nabla_{\overline{{X}}} g\big(\overline{{X}}^{*},\overline{{f}}^{*}\big) {+} \rho_{X} \big({L}_{\mathcal{G}}^{\top} \otimes {I}_{K_{\mathsf{TX}}} \big) \overline{{D}}_{X}^{*} = {0},$} \\
    \resizebox{.99\hsize}{!}{$\nabla_{\overline{{f}}} \mathcal{A} = {0} \implies \nabla_{\overline{{f}}} g\big(\overline{{X}}^{*},\overline{{f}}^{*}\big) {+} \rho_{f} \big({L}_{\mathcal{G}}^{\top} \otimes {I}_{M} \big) \overline{{D}}_{f}^{*} = {0}.$}
\end{gather*}

In addition, since $\lim_{k \rightarrow \infty} \big\|\big({L}_{\mathcal{G}} \otimes {I}_{M} \big) \overline{{f}}^{k+1}\big\|_{2} = 0$ and $\lim_{k \rightarrow \infty} \big\|\big({L}_{\mathcal{G}} \otimes {I}_{K_{\mathsf{TX}}} \big) \overline{{X}}^{k+1}\big\|_{F} = 0$ (c.f., Result~$1$), convergence to stationary points follows.

\textbf{Result~$3$: Sublinear Convergence Rate}

To derive convergence rates, we bound the optimality gap defined in the theorem statement. Per Result~$1$, for large enough $\xi > 0$, the optimality gap is upper bounded as
\begin{equation*}
    \resizebox{.99\hsize}{!}{$Q\big(\overline{{X}}^{k+1},\overline{{f}}^{k+1},\overline{{D}}_{X}^{k},\overline{{D}}_{f}^{k} \big) \leq \xi \big\| \overline{{X}}^{k+1} {-} \overline{{X}}^{k} \big\|_{F}^{2} {+} \xi \big\| \overline{{f}}^{k+1} {-} \overline{{f}}^{k} \big\|_{F}^{2}.$}
\end{equation*} 

Moreover, for some large enough $\nu > 0$, Result~$1$ implies
\begin{multline*}
    \mathcal{A}\big(\overline{{X}}^{k+1},\overline{{f}}^{k+1},\overline{{D}}_{X}^{k},\overline{{D}}_{f}^{k}\big) {-} \mathcal{A}\big(\overline{{X}}^{k},\overline{{f}}^{k},\overline{{D}}_{X}^{k},\overline{{D}}_{f}^{k}\big) \\ \leq {-}\nu \big\| \overline{{X}}^{k+1} {-} \overline{{X}}^{k} \big\|_{F}^{2} {-} \nu \big\| \overline{{f}}^{k+1} {-} \overline{{f}}^{k} \big\|_{F}^{2}.
\end{multline*}

On averaging over indices $k = 1, \cdots, K$, we obtain
\begin{multline*}
    \frac{1}{K} \sum\limits_{k = 1}^{K} Q\big(\overline{{X}}^{k+1},\overline{{f}}^{k+1},\overline{{D}}_{X}^{k},\overline{{D}}_{f}^{k} \big) \leq \\ \frac{\xi}{K \nu} \big(\mathcal{A}\big(\overline{{X}}^{1},\overline{{f}}^{1},\overline{{D}}_{X}^{1},\overline{{D}}_{f}^{1}\big) {-} \mathcal{A}\big(\overline{{X}}^{K+1},\overline{{f}}^{K+1},\overline{{D}}_{X}^{1},\overline{{D}}_{f}^{1}\big) \big).
\end{multline*}

In other words, the convergence in terms of the optimality gap is sublinear with rate $O\big({1}/{K}\big)$.
\end{proof}

\begin{remark}
To the best of our knowledge, we are not aware of any existing convergence analysis for ADMM that is directly applicable to our problem, which is distributed, integer constrained, bi-convex, and non-separable. On the one hand, in~\citep{Yin2013}, the authors consider the \emph{centralized} bi-convex and non-separable function class and show sublinear convergence assuming the objective function satisfies the Kurdyka–\L ojasiewicz (KL) inequality. On the other hand, in~\citep{Razaviyayn2016}, the authors consider the distributed non-convex but \emph{separable} function class and show convergence to stationary points. Neither analysis applies to our problem. However, in the absence of integer constraints, our problem is closely related to matrix factorization (MF). Hence, we have adapted prior convergence analysis for MF~\citep{Hajinezhad2016,Hajinezhad2018,Hong2017} to our problem. Since our problem is not identical to MF, our results have certain minor differences. Specifically, convergence to stationary points for MF only holds under certain regimes on the ADMM penalty parameters (e.g., $\rho > 1)$. No such conditions are required in our results.
\end{remark}

\begin{remark}
We remark that the applicability of ADMM to bi-convex problems is well-known from~\citep{Boyd2011}. Since strong duality does not hold in bi-convex problems, we can only demonstrate approximate convergence to KKT points. To do so, we have used the optimality gap function from~\citep{Hajinezhad2016,Hajinezhad2018,Hong2017} since it quantifies both first-order optimality conditions and the consensus error.
\end{remark}

We now characterize the algorithm complexity for DSISD.
\\
\begin{corollary}
    \label{corr:1}
    Let the same assumptions as in Theorem~\ref{thm:1} hold, and let $\epsilon > 0$ be a desired accuracy. Then, the algorithm complexity for DSISD is $O\big(2 |\mathcal{E}| \big(M {+} K_{\mathsf{TX}} (T{-}T_{1}) + \big(M{+}K_{\mathsf{RX}} T\big) \big({M N K_{\mathsf{RX}} T}\big)\big)/{\epsilon}\big)$.
\end{corollary}

\begin{proof}
The communication complexity is given by
\begin{equation*}
    O\big(2 |\mathcal{E}| \big(M {+} K_{\mathsf{TX}} (T{-}T_{1}) \big)/{\epsilon}\big),
\end{equation*}
which corresponds to the total number of real entries transferred across the network $\mathcal{G}$ over $K = O\big({1}/{\epsilon}\big)$ iterations as per the convergence rate in Theorem~\ref{thm:1}.

The computational complexity corresponds to the total cost of least-squares updates in every iteration and equals
\begin{equation*}
    O\big(N \big(M \big(K_{\mathsf{RX}}T\big)^{2} {+} M^{2} K_{\mathsf{RX}}T {+} K_{\mathsf{RX}} K_{\mathsf{TX}}^{2} {+} K_{\mathsf{TX}} K_{\mathsf{RX}}^{2} \big)/{\epsilon}\big).
\end{equation*}

Assuming the number of scatterers in the scene largely dominates the number of transmitting or receiving antennas, i.e., $M \gg K_{\mathsf{TX}},~K_{\mathsf{RX}}$, the algorithm complexity is given by the statement in the corollary.
\end{proof}

For comparison, consider decode-and-image. The communication complexity corresponds to every node transferring its $K_{\mathsf{RX}} \times T$ matrix of received symbols $Y_{T}^{(n)}$ to the central server, and thus equals $O(N K_{\mathsf{RX}} T)$. The computational complexity corresponds to least-squares updates over $N \times$ larger matrices, and equals $O(K_{\mathsf{RX}} N T M (M + K_{\mathsf{RX}} N T))$. Hence, DSISD has $O(N)$ smaller computational complexity compared to decode-and-image.

In the next section, we numerically evaluate the performance of DSISD and compare it with decode-and-image.

\begin{figure}[t]
    \centering
    \begin{subfigure}{.5\linewidth}
    \centering
    \includegraphics[page=1,width=0.95\linewidth]{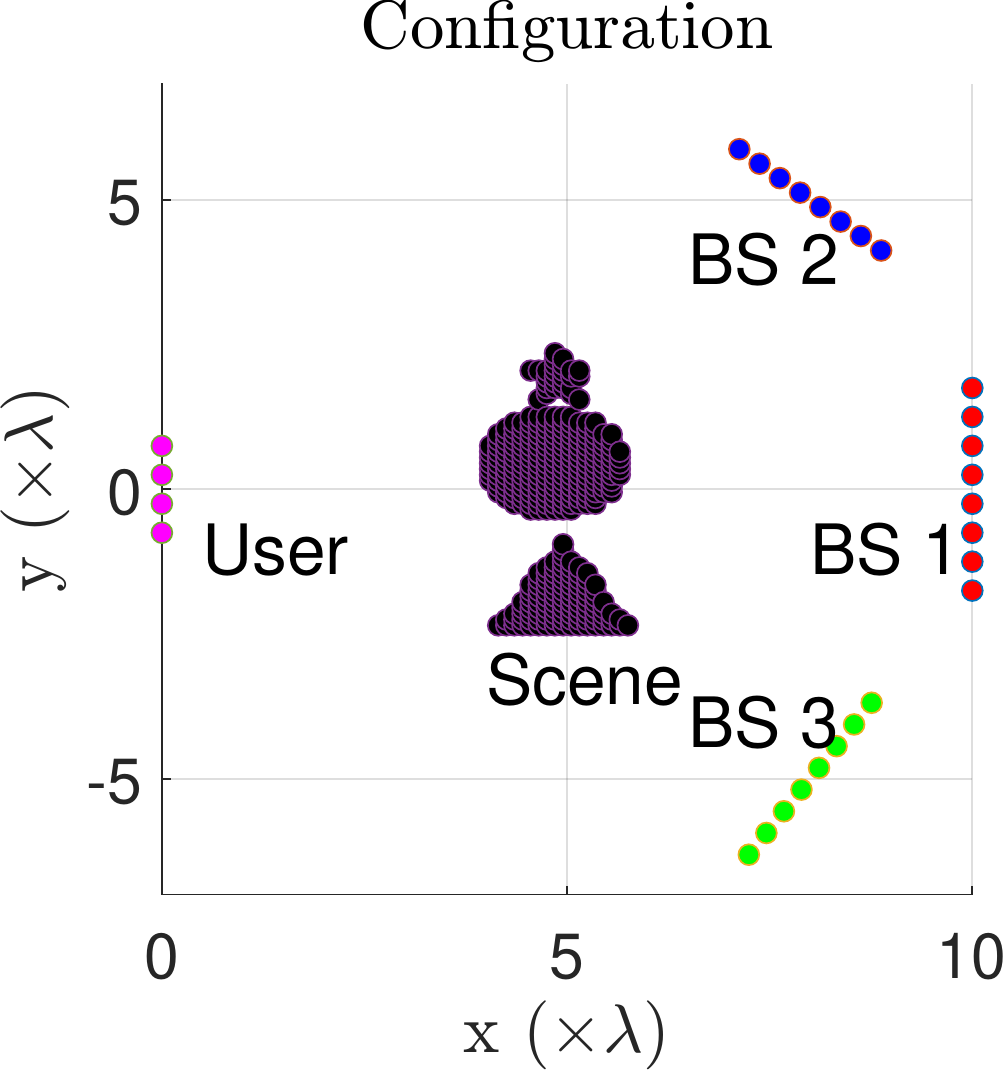}
    \caption{}
    \label{fig:plts_a}
    \end{subfigure}%
    \begin{subfigure}{.5\linewidth}
    \centering
    \includegraphics[page=1,width=0.95\linewidth]{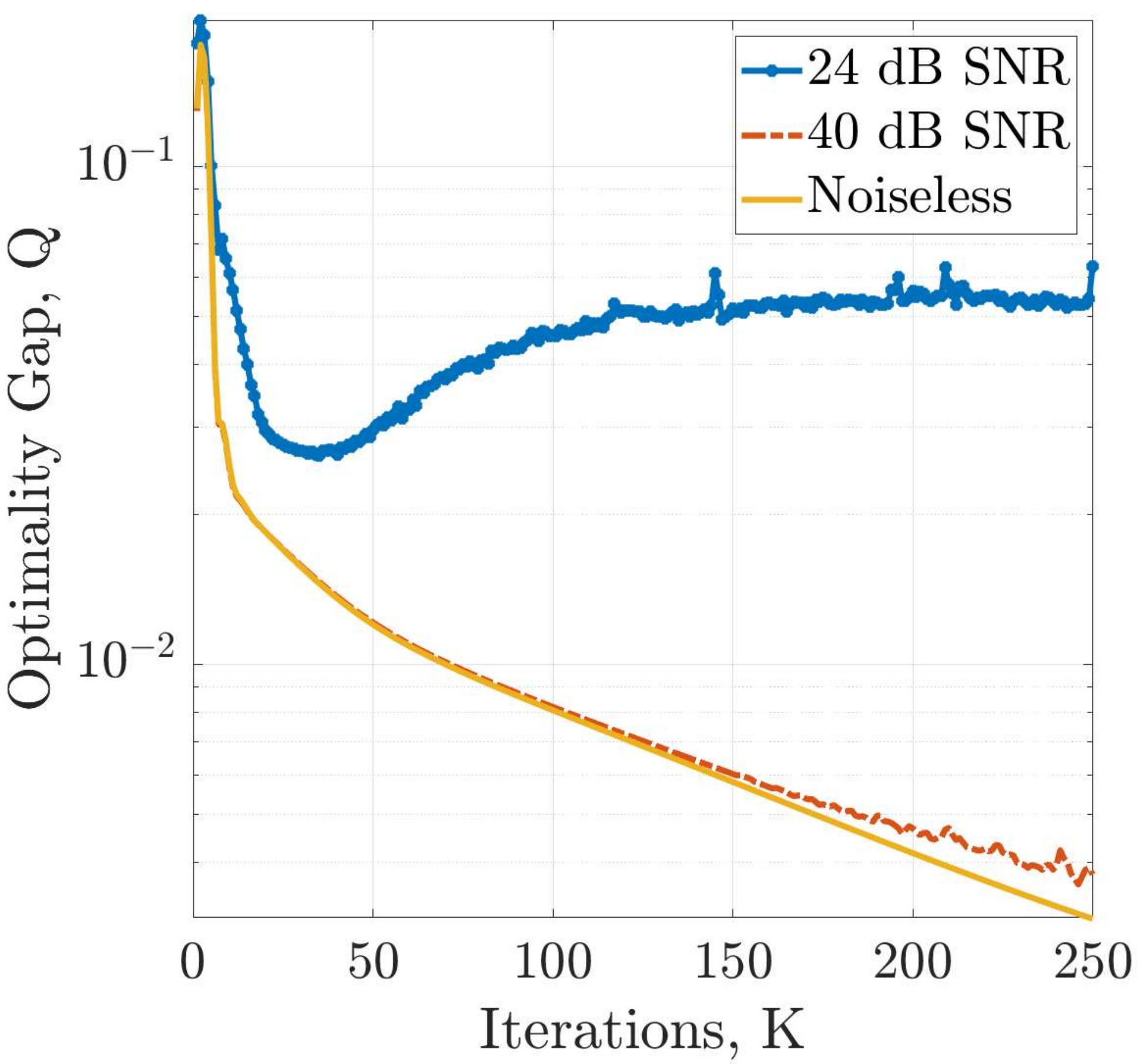}
    \caption{}
    \label{fig:plts_b}
    \end{subfigure}
    \caption{(a) Simulated configuration. (b) DSISD converges sublinearly in terms of optimality gap $Q$ for high receive signal-to-noise ratio (SNR) values.}
    \label{fig:plts1}
\end{figure}

\section{Numerical Evaluation}
\label{sec:num_eval}

\begin{figure*}[htbp]
    \begin{subfigure}{.3\linewidth}
    \centering
    \includegraphics[page=1,width=0.9\linewidth]{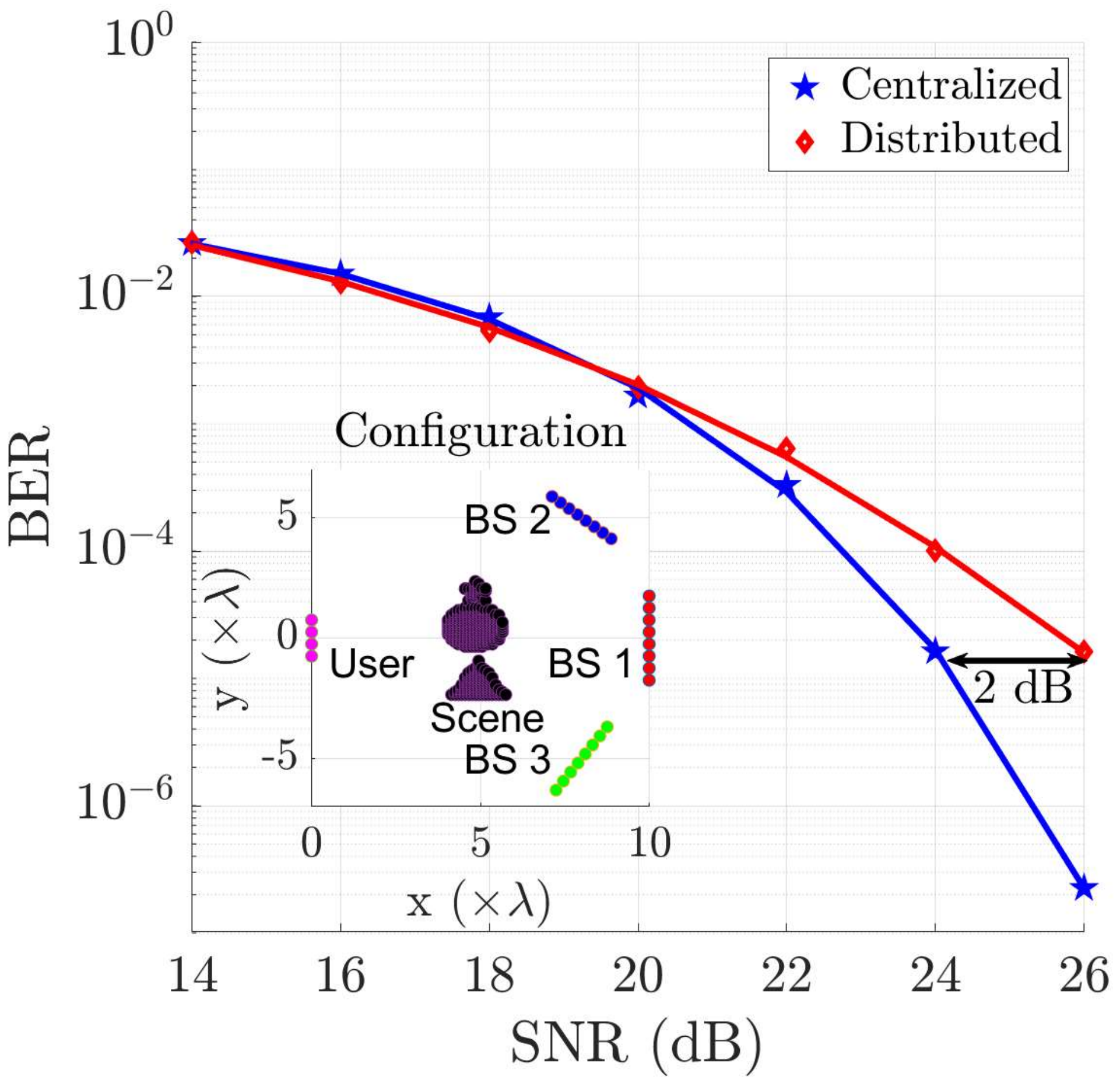}
    \caption{}
    \label{fig:plts_c}
    \end{subfigure}
    \vspace{1cm}
    \begin{subfigure}{.4\linewidth}
    \centering
    \includegraphics[page=1,width=0.7\linewidth,angle=270]{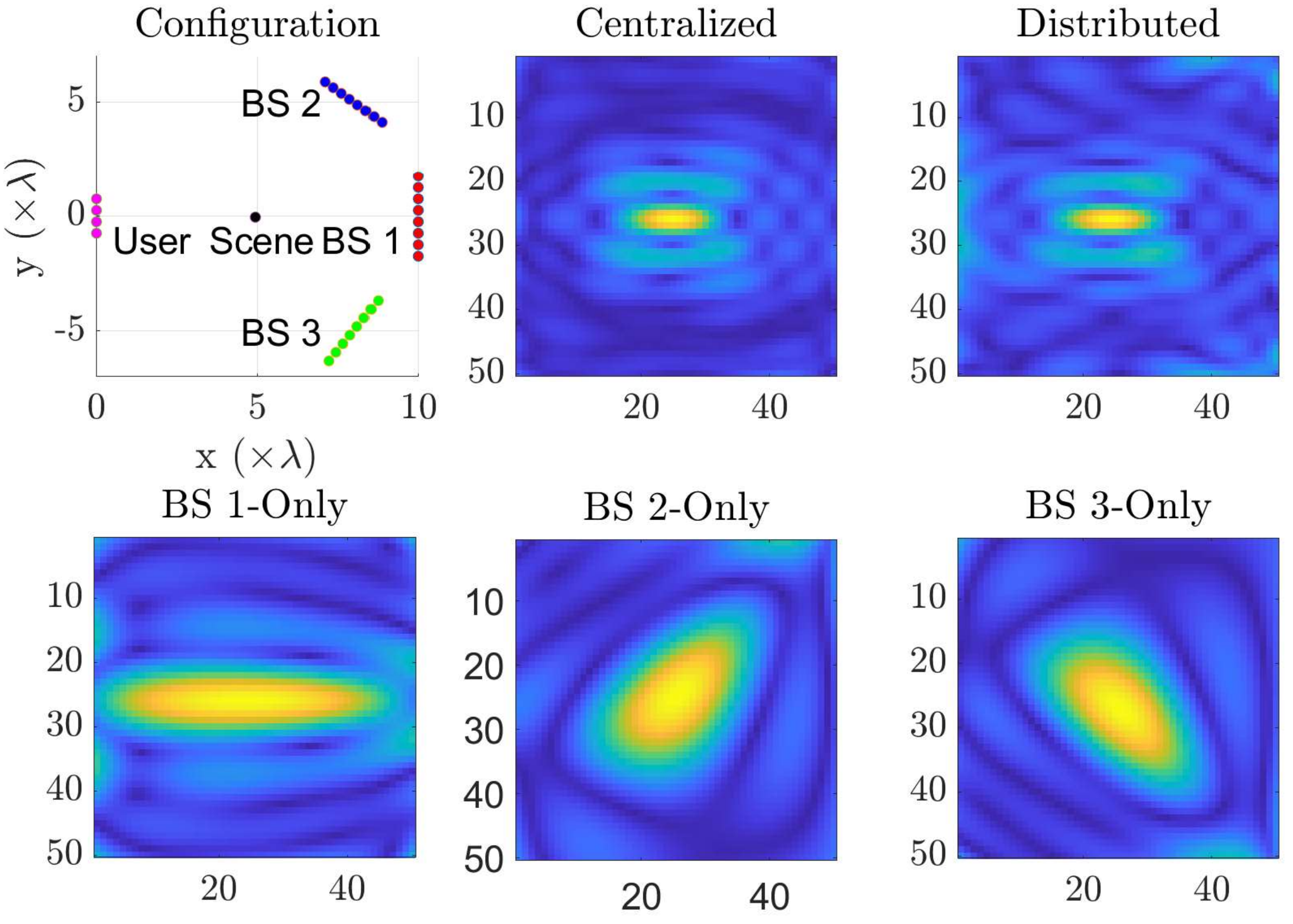}
    \caption{}
    \label{fig:plts_d}
    \end{subfigure}%
    \begin{subfigure}{.3\linewidth}
    \centering
    \includegraphics[page=1,width=0.9\linewidth]{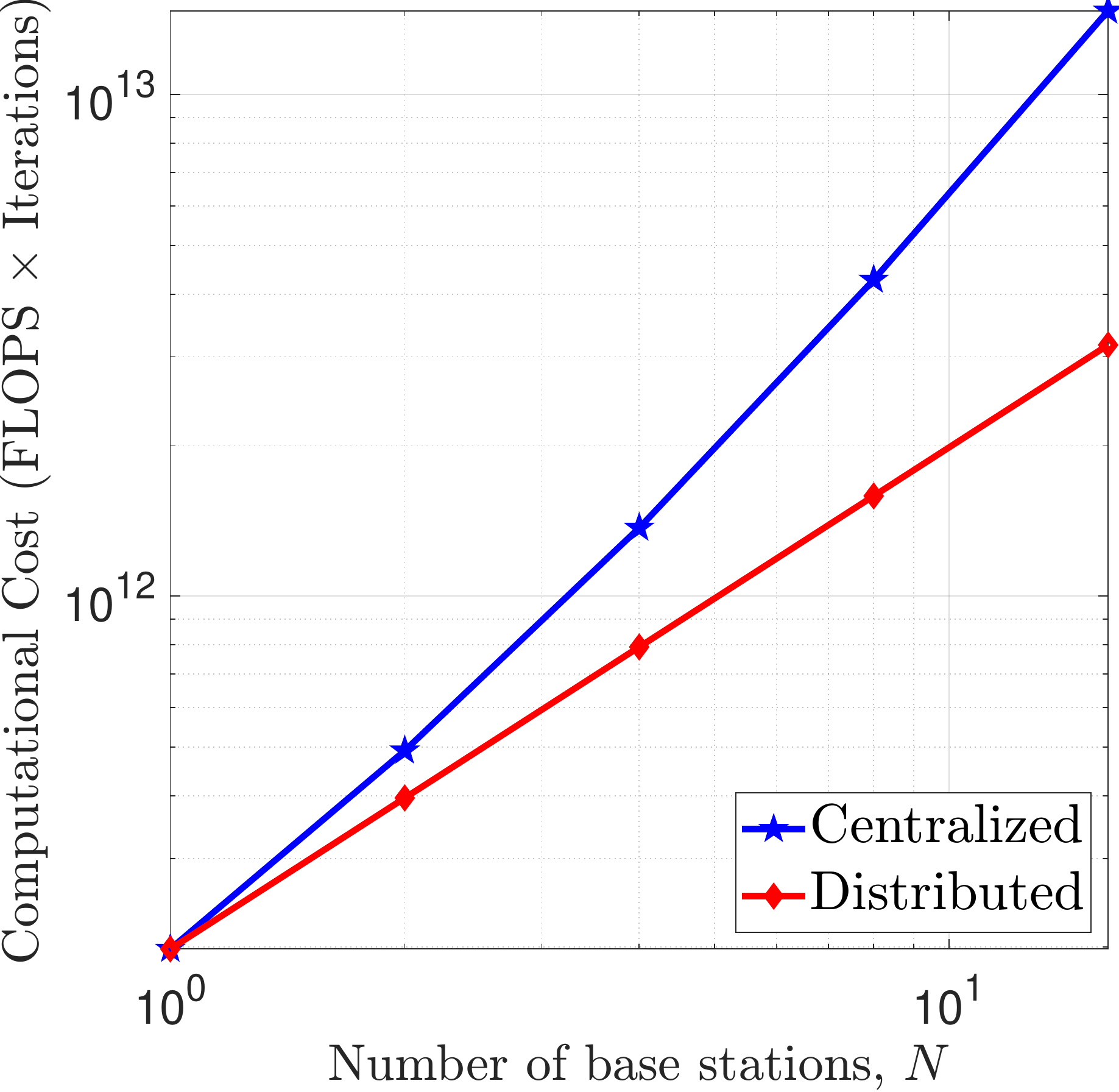}
    \caption{}
    \label{fig:plts_e}
    \end{subfigure}
    \caption{(a) DSISD achieves communication bit error rate (BER) within $2$ dB of centralized performance across a wide range of receive signal-to-noise ratio (SNR) values. (b) DSISD has similar imaging point spread functions (PSFs) as decode-and-image, with $3$ dB main lobe widths of $\lambda$ and $0.4 \lambda$ in range (x) and cross-range (y). Moreover, compared to local imaging performed at every base station, DSISD achieves resolution gains equivalent to jointly processing measurements across all $N = 3$ base stations. (c) DSISD has $O(N)$ smaller \text{computational} complexity (in total number of floating operations) compared to decode-and-image. All plots correspond to $K = 30$ iterations.}
    \label{fig:plts2}
\end{figure*}

We simulate the $2$D planar configuration shown in Fig.~\ref{fig:plts1}(\subref{fig:plts_a}) with $N = 3$ base stations receiving data transmissions from an uplink user in the $2.4$ GHz Wi-Fi band ($\lambda = 0.125$ m). We assume uncoded BPSK uplink transmissions, i.e., $\mathcal{X}_{T} = \left\{\pm 1\right\}^{K_{TX} \times T}$, with coherence interval $T = 100$ symbols. The uplink user is equipped with a uniform linear array (ULA) with $K_{\mathsf{TX}} = 4$ antennas. The base stations are each equipped with a ULA with $K_{\mathsf{RX}} = 8$ antennas, and are oriented at angles $0^{\circ}$, $-30^{\circ}$ and $45^{\circ}$ with respect to the uplink user's array.

For simplicity, we consider a star topology $\mathcal{G}$, where the base stations are all connected to a fusion node (not illustrated in Fig.~\ref{fig:plts1}(\subref{fig:plts_a})). The fusion node only processes the consensus information, i.e., local solutions ${f}_{k}^{(n)}$ and ${X}_{T-T_{1},k}^{(n)}$, exchanged by base stations, and does not process the received symbols ${Y}_{T}^{(n)}$ directly. The updates at the fusion node correspond to
\begin{align*}
    {\tilde{f}}_{k} &= \frac{{-}\sum\limits_{m} {L}_{\mathcal{G}}(N+1,m) {f}_{k-1}^{(m)} {+} {D}_{f,k-1}^{(N+1)}}{{L}_{\mathcal{G}}(N+1,N+1)}, \\
    {\tilde{X}}_{T-T_{1},k} &= \resizebox{.75\hsize}{!}{$\Pi_{\mathcal{X}_{T-T_{1}}}\Bigg(\frac{{-}\sum\limits_{m} {L}_{\mathcal{G}}(N+1,m) {X}_{T-T_{1},k-1}^{(m)} {+} {D}_{X,k-1}^{(N+1)}}{{L}_{\mathcal{G}}(N+1,N+1)}\Bigg)$},
\end{align*}
where the index $(N+1)$ denotes the fusion node. All remaining steps are identical to Algorithm~\ref{alg:admm_dist}, with $\overline{{f}}$ and $\overline{{X}}$ corresponding to the concatenation of local solutions over all $(N+1)$ nodes in the network.

First, we numerically evaluate the validity of our convergence analysis from Theorem~\ref{thm:1}. Fig.~\ref{fig:plts1}(\subref{fig:plts_b}) shows the optimality gap for DSISD at various receive signal-to-noise ratio (SNR) values. We observe that the sublinear $O({1}/{K})$ convergence predicted by Theorem~\ref{thm:1} holds in the high SNR regime. In future work, we shall refine our convergence analysis to incorporate the effect of SNR.

In Figs.~\ref{fig:plts2}(\subref{fig:plts_c}),~(\subref{fig:plts_d}) and~(\subref{fig:plts_e}), we show that DSISD achieves similar communication and imaging performance as decode-and-image, with $O(N)$ smaller \text{computational} complexity. In Fig.~\ref{fig:plts2}(\subref{fig:plts_c}), we show that DSISD achieves communication bit error rate (BER) within $2$ dB of decode-and-image. In Fig.~\ref{fig:plts2}(\subref{fig:plts_d}), we show that the imaging point spread functions (PSFs) for DSISD are similar to those for decode-and-image, with $3$ dB main lobe widths of $\lambda$ and $0.4 \lambda$ in range (x) and cross-range (y). Moreover, compared to local imaging performed at every base station, DSISD achieves resolution gains equivalent to jointly processing measurements across all $N = 3$ base stations. Finally, Fig.~\ref{fig:plts2}(\subref{fig:plts_e}) shows that the \text{computational} complexity (in total number of floating operations) for DSISD is $O(N)$ smaller compared to that of decode-and-image.

\section{Concluding Remarks}
\label{sec:conclusion}

We proposed DSISD, a provably convergent distributed algorithm based on consensus ADMM for simultaneous imaging and communication. We showed that DSISD achieves similar imaging and communication performance as centralized schemes with an order-wise reduction in \text{computational} complexity. We shall extend our convergence analysis to include the effects of integer constraints and SNR in future work. Moreover, we shall explore accelerated variants of DSISD with faster convergence rates. 


\bibliography{ifacconf}
                                                   


\end{document}